\documentstyle[12pt,psfig,cite]{article}
\begin{document}
\renewcommand{\thefootnote}{\fnsymbol{footnote}}
\begin{titlepage}
\begin{flushright}
KEK-TH-540 \\
NWU970901 \\
September 1997 \\
\end{flushright}
%
%\vspace*{10mm}
%\parskip 0.5cm
\huge
\begin{center}
%%%%%%%%%%%%%%%%%%%%%%%%%%%%%%%%%%%%%%%%%%%%%%%%%%%%%%%%%%%%%%%
Grand-canonical simulation of \\ two-dimensional simplicial gravity
%%%%%%%%%%%%%%%%%%%%%%%%%%%%%%%%%%%%%%%%%%%%%%%%%%%%%%%%%%%%%%%
\end{center}
\vspace{5mm}
\large
\begin{center}
S.Oda \footnote[2]{E-mail address: oda@phys.nara-wu.ac.jp},
N.Tsuda %$^{\dag}$ 
\footnote[3]{E-mail address: ntsuda@theory.kek.jp}
and 
T.Yukawa {\footnotesize $^{\S,\ddag}$}
\footnote[0]{\hspace{-1.5mm}{\tiny $^{\S}$}E-mail address: yukawa@theory.kek.jp}
\end{center}

\normalsize

\begin{center}
$^{\dag}$
Nara Women's University \\
Nara, Nara 630, Japan
\vspace{0.5cm}

$^{\ddag}$
High Energy Accelerator Research Organization (KEK) \\
       Tsukuba, Ibaraki 305, Japan
\vspace{0.5cm}

$^{\S}$
Coordination Center for Research and Education, \\
The Graduate University for Advanced Studies, \\
Hayama-cho, Miura-gun, Kanagawa 240-01, Japan 
\end{center}
\vspace{0.5cm}
%\parindent 5cm
%%%%%%%%%%%%%%%%%%%%%%
\begin{abstract}
%%%%%%%%%%%%%%%%%%%%%%
\noindent
The string susceptibility exponents of dynamically triangulated
2-dimensional surfaces with various topologies, such as a sphere,
torus and double-torus, were calculated by the grand-canonical
Monte Carlo method.
These simulations were made for surfaces coupled to $d$-Ising spins
($d$=0,1,2,3,5).
In each simulation the area of surface was constrained 
to within 1000 to 3000 of triangles,
while maintaining the detailed-balance condition.
The numerical results show excellent agreement with theoretical predictions
as long as $d \leq 2$.

\end{abstract}
\end{titlepage}
  
%%%%%%%%%%%%%%%%%%%%%%%%%%%%%%%%%%%%
\section{Introduction}
%%%%%%%%%%%%%%%%%%%%%%%%%%%%%%%%%%%%

In 2-dimensional quantum gravity
one of the most fundamental physical (universal) quantities is
the string susceptibility exponent ($\gamma$)
which characterizes the partition function
of surfaces with a fixed topology through
\begin{equation}
Z(A) \approx A^{\gamma-3} e^{\lambda_c A}
\end{equation}
for the large-$A$ limit,
where $A$ is the 2-dimensional volume (area),
and $\lambda_c$ is the cosmological constant.
The string susceptibility exponent is known to be given by the formula
\begin{equation}
\gamma = \frac{1-h}{12} 
         \left[ c-25 - \sqrt{(25-c)(1-c)} \right] + 2    \label{eq:gamma}
\end{equation}
in the continuum framework \cite{Liouville}
as well as the matrix model \cite{Matrix_model},
where $h$ is the number of handles of 
the 2-dimensional surface,
and $c$ represents the central charge of matter fields coupled to the surface.
Eq.(\ref{eq:gamma}) is expected to hold for a discretized 
2-dimensional surface with arbitral $h$ and $c \leq 1$, 
when the dynamical triangulation recipe \cite{DT} is 
adopted to generate random surfaces.

In order to measure the string susceptibility exponent with any topology, 
we introduced the grand-canonical Monte Carlo method. 
Among quantities of interest the string susceptibility exponent of 
a spherical surface
has been extensively calculated numerically by various methods
for examining the theoretical prediction, and a numerical method as well.
The MINBU (minimum neck baby universe) algorithm \cite{MINBU_sphere} 
and the grand-canonical Monte Carlo methods \cite{Grand_Canonical}
have been successful as long as the central charge is less than one
with sphere topology.
However, in the MINBU algorithm 
measuring $\gamma$ for a surface with higher genus 
is more complicated than for a sphere,
because the minimum neck baby universe must not be topologically non-trivial,
and the baby universe distribution describing 
two topologically different $\gamma$ 
must be separated independently \cite{MINBU_torus}.
Also, this algorithm is subject to rather large finite-size effects,
because the baby universes are dominated by small sizes,
even if the mother universe is large.
It cannot give reliable values for $c \neq 0$ unless we know the finite-size 
correction accurately \cite{Finite_size_effects}.
On the other hand, in the grand-canonical ensemble methods
the finite-size effect enters through the total area, 
and becomes controllable in reasonably large size simulations.
In this study we generated a sufficiently large size surface with a wider region
of area compared with the previous studies, which is essential to determine
$\lambda_c$ accurately.

Eq.(\ref{eq:gamma}) shows that $\gamma$ becomes complex 
for $1<c<25$ ($h\neq1$),
which is clearly unphysical and suggests
that the nature of the surface will change significantly
as $c$ moves beyond unity.
It is generally believed that this ``transition'' reflects the tachyonic 
instability of a bosonic string for $c>1$.
In a numerical simulation by dynamical triangulation 
the partition function is kept real for any value of $c$,
and the instability is considered to appear along with the collapse 
of the surface to branched polymers.
The string susceptibility exponent of the branched polymer 
($\gamma_{BP}$) depends on the model.
In fact, in ref.\cite{Gamma_BP_0.33_&_genus,Gamma_BP_0.33}
$\gamma_{BP}$ for a sphere is evaluated as $\frac{1}{3}$,
while in ref.\cite{Gamma_BP_0.33_&_genus,Gamma_BP_genus}
they calculated $\gamma_{BP}$ with any topology:
\begin{equation}
\gamma_{BP} = \frac{1}{2} + \frac{3}{2} h.  \label{eq:branch}
\end{equation}
In the case of a torus ($h=1$) the effect of the tachyon cannot be seen in the
string susceptibility exponent, 
because both eq.(\ref{eq:gamma}) and eq.(\ref{eq:branch})
give the  same value of $\gamma = 2$ for any $c$.

In order to check the theoretical expectations 
(eqs.(\ref{eq:gamma}) and (\ref{eq:branch})), 
we need to calculate the string susceptibility exponent for a surface 
with any topology.

We describe details of our method in Section \ref{sec:method},
and our numerical results in Section \ref{sec:results}.
Section \ref{sec:discussion} is devoted to a summary and discussions.

%%%%%%%%%%%%%%%%%%%%%%%%%%%%%%%%%%%%
\section{The method} \label{sec:method}
%%%%%%%%%%%%%%%%%%%%%%%%%%%%%%%%%%%%
%%%%%%%%%%%%%%%%%%%%%%%%%%%%%%%%%%%%
\subsection{How to evaluate $\gamma$}
%%%%%%%%%%%%%%%%%%%%%%%%%%%%%%%%%%%%
The grand-canonical partition function with a fixed topology is defined by
\begin{equation}
Z(\lambda) = \sum_{T,N_2} W(T,N_2) e^{-\lambda N_2} Z_{\sigma},
                                                      \label{eq:Z_lambda}
\end{equation}
where
\begin{equation}
Z_{\sigma} = \exp \left\{ \beta \sum_{\mu=1}^d \sum_{<ij>} 
                         \sigma_i^{\mu} \sigma_j^{\mu}         \right\}. 
                                                      \label{eq:Z_sigma}
\end{equation}
In eqs.(\ref{eq:Z_lambda}) and (\ref{eq:Z_sigma})
$T$ refers to a specific triangulation, $W(T,N_2)$ is the symmetry factor, 
$N_i$ is the number of $i$-simplices 
({\it i.e.} $N_0$, $N_1$ and $N_2$ are the number of vertices, links 
and triangles),
$\lambda$ is a parameter, and $\beta$ is a coupling strength of Ising spins.
There are $d$ kinds of Ising spins ($\sigma^{\mu}_i$),
and each Ising spin lives on a vertex of triangulated surface
(not on the dual lattice).
One kind of Ising spins carries $\frac{1}{2}$ central charge
at the critical point $\beta = \beta_c$,
and thus  the $d$ kinds of Ising spins correspond to $c=\frac{d}{2}$.

The partition function is written for the large-$N_2$ limit as 
\begin{equation}
Z(\lambda) \approx \sum_{N_2} N_2^{\gamma -3}e^{-(\lambda - \lambda_c) N_2}.
\end{equation}
Then, the probability of fixing a surface with $N_2$ triangles ($P(N_2)$) 
for large $N_2$ is approximately given by 
\begin{equation}
P(N_2) \propto N_2^{\gamma -3}e^{-(\lambda-\lambda_c)N_2}.  
                                                     \label{eq:distribution}
\end{equation}
From
\begin{equation}
\frac{\ln P(N_2)}{\ln N_2} = (\gamma -3) 
                       - (\lambda-\lambda_c)\frac{N_2}{\ln N_2} + O(1/\ln N_2),
                                                   \label{eq:log_distribution}
\end{equation}
if we tune the parameter $\lambda$  close to $\lambda_c$ 
within the precision 
\begin{equation}
| \lambda - \lambda_c | \ll \frac{\ln N_2}{N_2},
\end{equation}
we can evaluate the string susceptibility exponent 
by plotting $\ln P(N_2)$ as a function of $\ln N_2$.

The parameter $\lambda$ is determined in the following manner.
If $\lambda$ is chosen to be less (or greater) than $\lambda_c$, 
the system will expand (or shrink) exponentially 
in the grand-canonical simulation.
By requiring an exponential increase (or decrease) 
to disappear in the simulation 
we can find the parameter $\lambda$ precisely.
We then generate configurations dynamically
near to the critical point $\lambda_c$, constraining $N_2$ within a
upper- and a lower-bound ($N_2^{min} \leq N_2 \leq N_2^{max}$).

%%%%%%%%%%%%%%%%%%%%%%%%%%%%%%%%%%%%%%%%%%%%%%%%%%%%%%%%%%%%%%
\subsection{Procedure of the grand-canonical simulation}
%%%%%%%%%%%%%%%%%%%%%%%%%%%%%%%%%%%%%%%%%%%%%%%%%%%%%%%%%%%%%%
%%%%%%%%%%%%%%%%%% grand-canonical ensembles  %%%%%%%%%%%%%%%%
We generate dynamically triangulated surfaces 
using the grand-canonical Monte Carlo method.
In order to keep $N_2$ within the appropriate range
wide enough to extract $\gamma$ from eq.(\ref{eq:log_distribution}),
but not too large for the sake of the computing time,
we simply ignore all of the moves 
that make the number of triangles become less than 
$N_2^{min}$ or greater than $N_2^{max}$.
Here, we restrict ourself to graphs excluding ``tadpole'' and 
``self-energy'' in the dual lattice.
We note that in the grand-canonical method 
we must pay careful attention to satisfying the detailed valance,
because the number of states depends on the total number of triangles
and the coordination number of each vertex.

We first explain the method used to generate dynamically triangulated surfaces
in the case of pure gravity ($c=0$). 
%We describe it here in terms of the dual diagram.
Suppose that we wish to increase the number of real vertices 
from $N_0$ to $N_0+1$;
to do this we choose two dual links belonging to the same dual loop, 
and insert a ``propagator'' dual link joining the two dual links,
thus creating an additional dual loop.
For the inverse move, $N_0 \rightarrow N_0 -1$, we choose a dual link at random,
and remove it, 
resulting in two dual loops separated by the dual link 
merging into one dual loop.
The total number of possible increase (or decrease) moves 
from configuration $A$ are
\begin{equation}
n_A^{inc} = \sum_{i=1}^{N_0} \frac{q_i(q_i-1)}{2}, 
\end{equation}
\begin{equation}
n_A^{dec} = \sum_{i=1}^{N_0} \frac{q_i}{2},
\end{equation}
where $\{ q_i \}$ is the set of coordination numbers of configuration $A$.

In these moves the ergodic property is automatically satisfied.
However, we must treat the detailed valance carefully, 
since the total number of possible moves depends on 
$N_0$ and  $\{ q_i \}$.
We must multiply the prefactor to the Boltzmann weight  
in order to satisfy the detailed valance equation for the moves 
between configurations $A$ and $B$, given by
\begin{equation}
\frac{W_A}{n_A} P_{A \rightarrow B} =
\frac{W_B}{n_B} P_{B \rightarrow A},              \label{eq:detail_bal}
\end{equation}
where $P_{A \rightarrow B}$ is the transition probability from configurations
$A$ to $B$, $W_A$ is the Boltzmann factor, and
$n_A$ is the number of possible moves from configuration $A$.
For detailed arguments for eq.(\ref{eq:detail_bal}) on random lattices,
see ref.\cite{Detail_bal}.

For the case $d \neq 0$,
when a dual loop is separated by two dual loops,
two spins should be attached to the two dual loops,
which counts four spin states; 
in the inverse case one of two neighboring spins
should be taken out, which counts two spin states.
The combined number of operations and the Boltzmann factor are
\begin{equation}
n_A = \sum_{i=1}^{N_0}\left\{ \frac{q_i(q_i-1)}{2} \cdot 4^d 
                           + \frac{q_i}{2} \cdot 2^d          \right\},
\end{equation}
\begin{equation}
W_A = \exp \left\{ - \lambda N_2 + 
        \beta \sum_{\mu=1}^d \sum_{<ij>}\sigma_i^{\mu} \sigma_j^{\mu}      
                                                         \right\}.
\end{equation}
For the spin variables, Monte Carlo trials with Wolff's cluster algorithm 
\cite{Cluster} are performed after some geometrical moves.

%%%%%%%%%%%%%%%%%%%%%%%%%%%%%%%%%%%%%%%%%%%%%%%%%%%%%%%%%%%%%%%%%%%%%%%%
\section{Numerical results} \label{sec:results}
%%%%%%%%%%%%%%%%%%%%%%%%%%%%%%%%%%%%%%%%%%%%%%%%%%%%%%%%%%%%%%%%%%%%%%%%
\begin{table}
\caption{$Sphere$}
\begin{center}
\begin{tabular}{|l|lll|} \hline
h=0	& $\gamma$	& $\lambda_c$	& $\beta_c$	\\ \hline
c=0	& -0.517(27)	& 1.12467	& --------	\\ 
c=0.5 	& -0.345(27)	& 1.53716	& 0.227 	\\ 
c=1.0	& -0.009(26)	& 1.94848	& 0.226 	\\ 
c=1.5	&  0.112(27)	& 2.35843	& 0.225 	\\ 
c=2.5	&  0.238(26)	& 3.17513	& 0.223 	\\ \hline
\end{tabular}
\end{center}
\label{Tabel:0genus}
\end{table}
\begin{table}
\caption{$Torus$}
\begin{center}
\begin{tabular}{|l|lll|}       \hline
h=1	& $\gamma$	& $\lambda_c$	& $\beta_c$	\\ \hline
c=0	& 2.014(24)	& 1.12469	& --------	\\  	
c=0.5 	& 1.995(23)	& 1.52890	& 0.2163	\\ 
c=1.0	& 1.971(22)	& 1.93335	& 0.2163	\\ 
c=1.5	& 1.980(23)	& 2.33841	& 0.2163	\\ 
c=2.5	& 2.018(25)	& 3.14950	& 0.2163	\\ \hline
\end{tabular}
\end{center}
\label{Tabel:1genus}
\end{table}
\begin{table}
\caption{$Double-torus$}
\begin{center}
\begin{tabular}{|l|lll|}       \hline
h=2	& $\gamma$	& $\lambda_c$	& $\beta_c$	\\ \hline
c=0	& 4.510(27)	& 1.12471	& --------  	\\ 
c=0.5 	& 4.338(26)	& 1.52821	& 0.215  	\\ 
c=1.0	& 4.043(27)	& 1.93039	& 0.214 	\\ 
c=1.5	& 3.909(26)	& 2.33123	& 0.213 	\\ 
c=2.5	& 3.781(27)	& 3.12972	& 0.211 	\\ \hline
\end{tabular}
\end{center}
\label{Tabel:2genus}
\end{table}

We have measured the string susceptibility exponent
for various combinations of 
$c= \frac{d}{2} = 0,\frac{1}{2},1,\frac{3}{2},\frac{5}{2}$ 
and $h=0,1,2$.
The central charge and topology of the surface are kept fixed 
during each simulation.
When Ising spins are put on the surface,
we search the critical coupling strength ($\beta_c$) 
with the fixed number of triangles set at 3000 
before making a grand-canonical simulation.
We determine the critical coupling strength
from the peak of magnetic susceptibility 
for each combination of $c$ and $h$.
In the case of a torus the experimental values ($\beta_c$) were quite close to
the theoretical value ($\beta_c \approx 0.2163$),
and we used the exact value;
in the other cases, however, we used the experimental $\beta_c$.
We set the maximum number of triangles ($N_2^{max}$) to $3000$ and
the minimum one ($N_2^{min}$) to $1000$.
One sweep is defined by $2000 \times 2^d$ trial moves 
and 5 cluster operations of Ising spin.

\begin{figure}
\centerline{\psfig{file=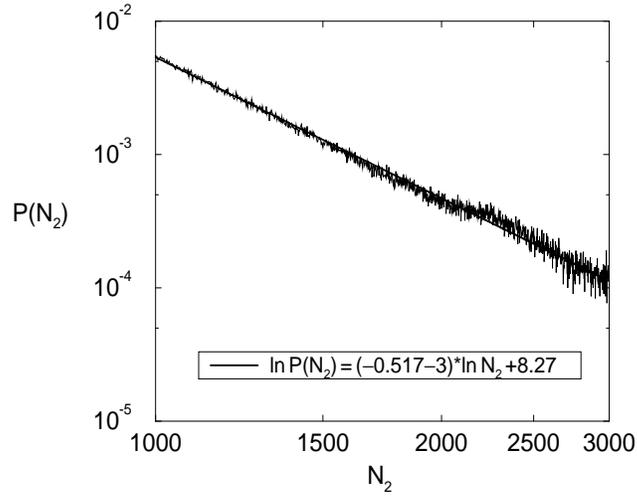,height=7cm,width=8cm}} 
\caption
{
Distribution ($P(N_2)$) 
for spherical pure gravity with log-log scales.
The parameter $\lambda$ is $1.12467$, 
and the string susceptibility exponent ($\gamma$) is $-0.517$.
}
\label{Fig:0genus0spin}
\end{figure}

In order to obtain $P(N_2)$, we measure the number of triangles ($N_2$) 
for $10^5$ times every 5 sweeps.
Plotting $\ln P(N_2)$ versus $\ln N_2$, such as Fig.\ref{Fig:0genus0spin},
we determine the string susceptibility exponent from its slope.
Our results are summarized in Tables 1, 2 and 3 
for a sphere, torus and double-torus, respectively.
%\ref{Table:0genus},\ref{Table:1genus},\ref{Table:2genus}.
Especially, in the pure gravity case we know the theoretical value,
\begin{equation}
\lambda_c = \frac{1}{2} \ln \frac{256}{27} \approx 1.12467,
\end{equation}
and our experimental $\lambda_c$ is very close to theoretical prediction.
The errors indicated in parentheses for the last two digits are only 
statistical errors estimated from the least-squares fits.
The systematic errors are due mainly to how close we can determine  $\lambda$
to $\lambda_c$.
In our simulation this systematic error is estimated to be around
$0.02 \sim 0.03$.

%%%%%%%%%%%%%%%%%%%%%%%%   0genus0spin   %%%%%%%%%%%%%%%%%%%%%%%%%%
%%%%%%%%%%%%%%%%%%%%%%%%%%%%%%%%%%%%%%%%%%%%%%%%%%%%%%%%%%%%%%%%%%%%
%%%%%%%%%%%%%%%%%%%%%%%%   Pure gravity   %%%%%%%%%%%%%%%%%%%%%%%%%%
\begin{figure}
\centerline{\psfig{file=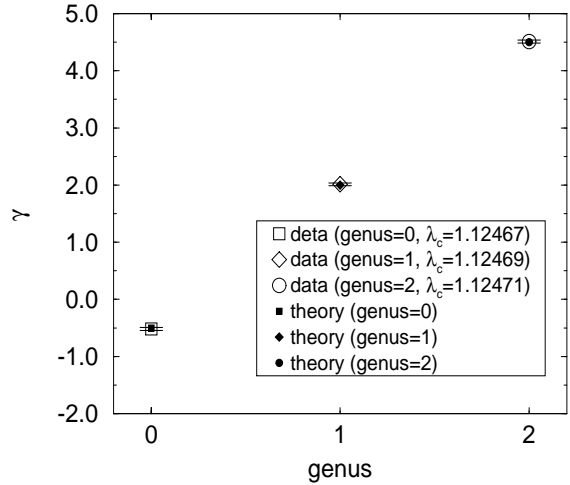,height=7cm,width=8cm}} 
\caption
{
String susceptibility exponents ($\gamma$) of 
a sphere, torus and double-torus
for pure gravity ($c=0$).
}
\label{Fig:PureGravity} 
\end{figure}
%%%%%%%%%%%%%%%%%%%%%%%%   Sphere   %%%%%%%%%%%%%%%%%%%%%%%%%%
\begin{figure}
\centerline{\psfig{file=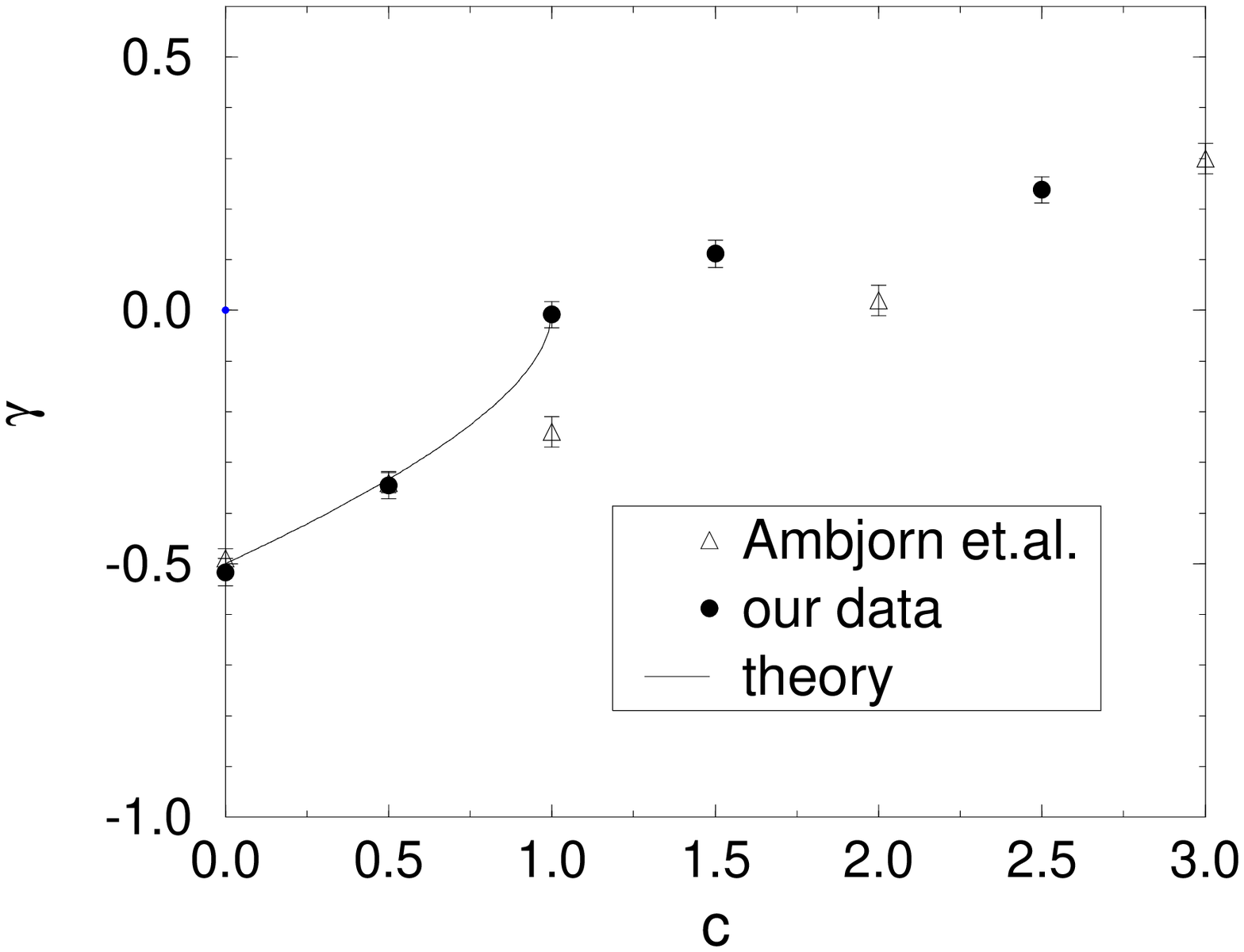,height=7cm,width=8cm}} 
\caption
{
String susceptibility exponents ($\gamma$) of
a sphere with various central charges ($c$).
Solid line represents eq.(2) with $h=0$.
}
\label{Fig:Sphere}
\end{figure}
%%%%%%%%%%%%%%%%%%%%%%%%%%%%%%%%%%%%%%%%%%%%%%%%%%%%%%%%%%%%%%%%%%%%
%%%%%%%%%%%%%%%%%%%%%%%%   Torus   %%%%%%%%%%%%%%%%%%%%%%%%%%
\begin{figure}
\centerline{\psfig{file=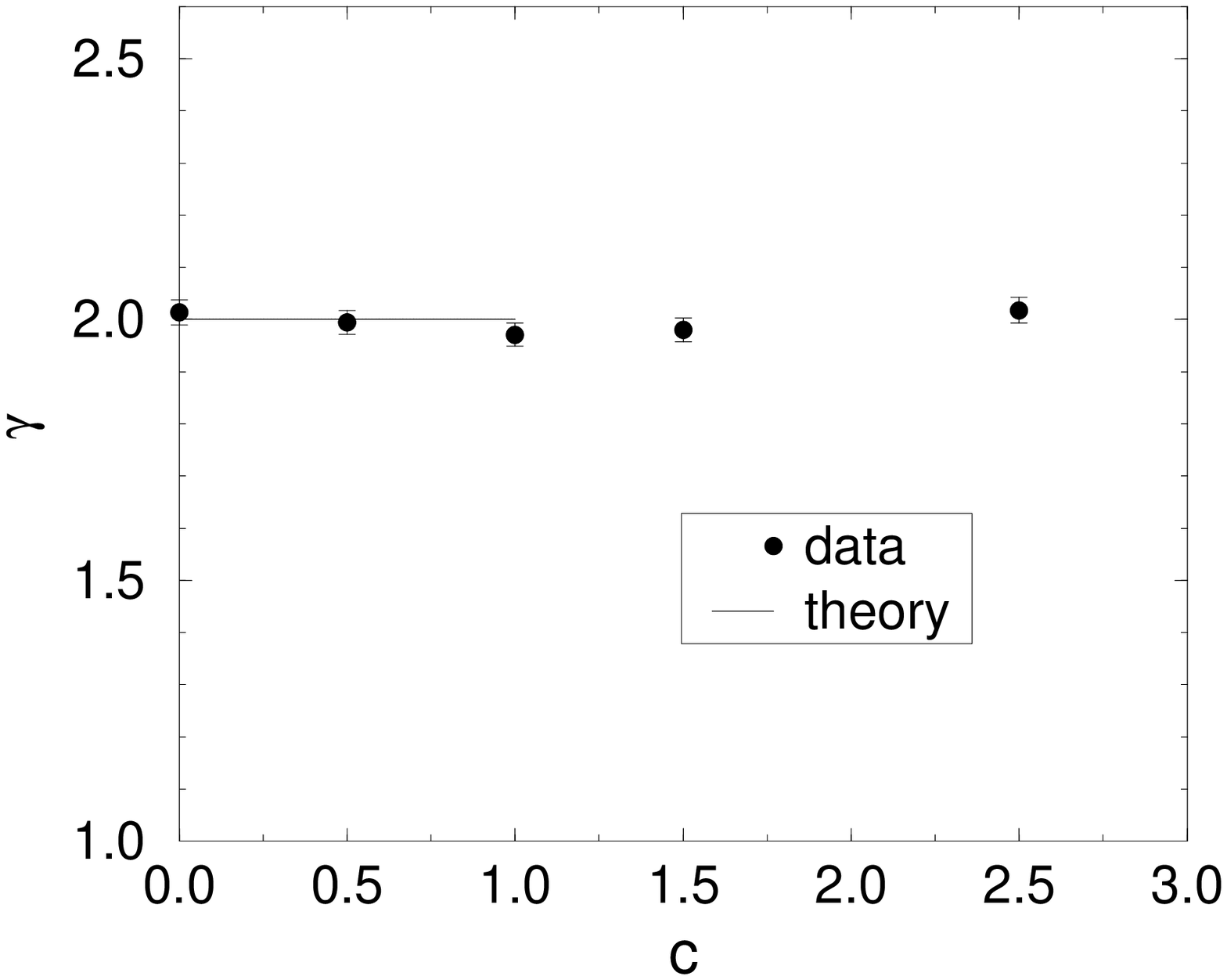,height=7cm,width=8cm}} 
\caption
{
String susceptibility exponents ($\gamma$) of
a torus with various central charges ($c$).
Solid line represents eq.(2) with $h=1$.
}
\label{Fig:Torus}
\end{figure}
%%%%%%%%%%%%%%%%%%%%%%%%%%%%%%%%%%%%%%%%%%%%%%%%%%%%%%%%%%%%%%%%%%%%
%%%%%%%%%%%%%%%%%%%%%%%%   Double-Torus   %%%%%%%%%%%%%%%%%%%%%%%%%%
\begin{figure}
\centerline{\psfig{file=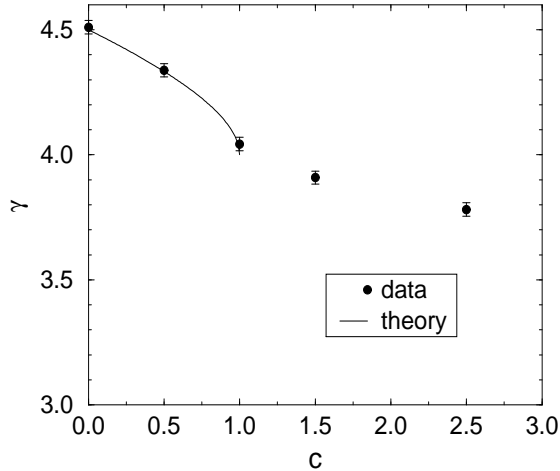,height=7cm,width=8cm}} 
\caption
{
String susceptibility exponents ($\gamma$) of
a double-torus with various central charges ($c$).
Solid line represents eq.(2) with $h=2$.
}
\label{Fig:Double-Torus}
\end{figure}
%%%%%%%%%%%%%%%%%%%%%%%%%%%%%%%%%%%%%%%%%%%%%%%%%%%%%%%%%%%%%%%%%%%%

%%%%%%%%%%%%%%%%%%%%%%%%%%%%%%%%%%%%%%%%%%%%%%%%%%%%%%%%%%%%%%%%%%%%%%%%
\section{Summary and discussion} \label{sec:discussion}
%%%%%%%%%%%%%%%%%%%%%%%%%%%%%%%%%%%%%%%%%%%%%%%%%%%%%%%%%%%%%%%%%%%%%%%%
We calculated the string susceptibility exponents
of 2-dimensional triangulated surfaces
coupled to $d$-Ising spins ($d$=0,1,2,3,5)
with various topologies ($h$=0,1,2)
using the grand-canonical Monte Carlo method.
One of the advantages of our method is the ability to treat various topologies.
Fig.\ref{Fig:PureGravity} shows good agreements to the theory
for not only a sphere, 
but also for a torus and a double-torus of the pure gravity.
Furthermore, in Figs.\ref{Fig:Sphere},\ref{Fig:Torus} and \ref{Fig:Double-Torus}
the string susceptibility exponents for various topologies ($h=0,1,2$)
are reproduced very closely to the theoretical predictions as long as $c \leq 1$.
Especially for $c=1$ of the spherical surface 
we give the theoretically expected $\gamma$.
The log correction in the partition function is one order smaller than the 
leading term in our simulation for $N_2$, being on the order $10^3$,
which proves that our method is free from a finite-size correction,
compared to the MINBU algorithm.

We are especially interested in the behavior of $\gamma$ for $c>1$.
In the case of a torus ($h=1$)
we cannot observe the effect of the central charge 
in Fig.\ref{Fig:Torus}, as expected from the theoretical prediction in 
eq.(\ref{eq:gamma}) and eq.(\ref{eq:branch}).
In the spherical case ($h=0$) $\gamma$ becomes positive and increases slowly,
while in the double-torus ($h=2$) it seems to decrease slowly.
Although 2-dimensional surfaces seem to make a transition 
to the branched polymer phase gradually as $c>1$,
based on these figures,
we cannot conclude for certain
what kind of branched polymers were obtained in our simulations.  
It is preferable to find other quantities to observe the branched polymer 
phase more clearly.
For example, in the case of a higher genus the moduli of the surface is
one of the most important quantities in order to distinguish 
the branched polymer\cite{Cmplx_Struc}.

\begin{center}
{\Large -Acknowledgements-}
\end{center}
It is a pleasure to thank H.Kawai for many helpful discussions.
Two of the authors(N.T. and S.O.) are supported by Research Fellowships of the 
Japan Society for the Promotion of Science for Young Scientists.

%%%%%%%%%%%%%% bibliography %%%%%%%%%%%%%


\begin{thebibliography}{99}
%
%%%%%%%%%%%%%%%%%%%% Liouville %%%%%%%%%%%%%%%%%
\bibitem{Liouville} 
%%\bibitem{KPZ} 
V.G.Knizhnik, A.M.Polyakov and A.B.Zamolodchikov, 
Mod.Phys.Lett A, Vol.3 (1988) 819; 
%%\bibitem{Dis_Kaw} 
J.Distler and H.Kawai, 
Nucl.Phys. B321 (1989) 509;
%%\bibitem{Dav} 
F.David, 
Mod.Phys.Lett. A3 (1988) 1651.
%%%%%%%%%%%%%% A matrix model %%%%%%%%%%%%
\bibitem{Matrix_model}
%%\bibitem{Bre_Kaz} 
E.Br\'{e}zin and V.Kazakov, 
Phys.Lett. B236 (1990) 144;
%%\bibitem{Dou_She} 
M.Douglas and S.Shenker, 
Nucl.Phys. B335 (1990) 635; 
%%\bibitem{Gro_Mig} 
D.J.Gross and A.A.Migdal, 
Phys.Rev.Lett. 64 (1990) 717.
%%%%%%%%%%%%%%%%%%%%% DT %%%%%%%%%%%%%%%%%%%
\bibitem{DT}
%\bibitem{Kaz} 
V.A.Kazakov, I.K.Kostov and A.A.Migdal, 
Phys.Lett. B157 (1985) 295; 
%\bibitem{Amb_Dur_Fre} 
J.Ambj\o rn, B.Durhuus and J.Fr$\ddot{o}$hlich, 
Nucl.Phys.B257[FS14] (1985) 433;
%\bibitem{Davi} 
F.David, 
Nucl.Phys. B257[FS14] (1985) 543.
%%%%%%%%%%%%%%%%%%%%%% MINBU sphere %%%%%%%%%%%%%%%%%%%%%%%%%%%%%
\bibitem{MINBU_sphere}
S.Jain and S.D.Mathur, Phys.Lett.B286 (1992) 239;
J.Ambj\o rn, S.Jain and G.Thorleifsson, Phys.Lett.B307 (1993) 34;
J.Ambj\o rn and G.Thorleifsson, Phys.Lett.B323 (1994) 7.
%%%%%%%%%%%%%%%%%%%%% Grand Canonical Simulation %%%%%%%%%%%%%%%%%%%
\bibitem{Grand_Canonical}
J.Jurkiewicz, A.Krzywicki and B.Petersson, Phys.Lett. B177 (1986) 89;
J.P.Kownacki and A.Krzywicki, Phys.Rev. D50 (1994) 5329.
%%%%%%%%%%%%%%%%%%%%%% MINBU torus %%%%%%%%%%%%%%%%%%%%%%%%%%%%%%
\bibitem{MINBU_torus}
G.Thorleifsson and S.Catterall, Nucl.Phys.B461 (1996) 350.
%%%%%%%%%%%%%%%%%%%%%% Finite size effects %%%%%%%%%%%%%%%%%%%%%%
\bibitem{Finite_size_effects}
N.D.Hari Dass, B.E.Hanlon and T.Yukawa, Phys.Lett.B368 (1996) 55.
%%%%%%%%%%%%%%%%%%%%%% Gamma_BP 1/3 & genus %%%%%%%%%%%%%%%%%%%%%
\bibitem{Gamma_BP_0.33_&_genus}
B.Dorhuus, Nucl.Phys.B426 (1994) 203.
%%%%%%%%%%%%%%%%%%%%%% Gamma_BP 1/3 %%%%%%%%%%%%%%%%%%%%%%%%%%%%%
\bibitem{Gamma_BP_0.33}
J.Ambj\o rn, B.Dorhuus and T.Jonsson, Mod.Phys.Lett.A, Vol.9 (1994)1221.
%%%%%%%%%%%%%%%%%%%%%% Gamma_BP genus %%%%%%%%%%%%%%%%%%%%%%%%%%%
\bibitem{Gamma_BP_genus}
J.Jurkiewicz and A.Krzywicki,  Phys.Lett.B392 (1997) 291.
%%%%%%%%%%%%%%%%%%%%%% Detailed balance %%%%%%%%%%%%%%%%%%%%%%%%%
\bibitem{Detail_bal}
N.Tsuda , A.Fujitsu and T.Yukawa,
Comp.Phys.Comm. 87 (1995) 372.
%%%%%%%%%%%%%%%%%%%%%% Cluster %%%%%%%%%%%%%%%%%%%%%%%%%%%%%%%%%%
\bibitem{Cluster}
U.Wolff, Phys.Rev.Lett.62 (1989) 361.
%%%%%%%%%%%%%%%%%%%%%% Complex structures %%%%%%%%%%%%%%%%%%%%%%%
\bibitem{Cmplx_Struc}
H.Kawai, N.Tsuda and T.Yukawa, Phys.Lett.B351 (1995) 162,
Nucl.Phys.B (Proc.Suppl.) 47 (1996) 653, {\it Frontiers in Quantum 
Field Theory}, Ed. by H.Itoyama et.al., (World Scientific, 1995).
%%%%%%%%%%%%%% Fractal structures of QG surfaces %%%%%%%%%%%%%%
%\bibitem{Fractal} 
%H.Kawai and M.Ninomiya, 
%Nucl.Phys.B336 (1990) 115;
%%\bibitem{Agi_Mig}
%M.E.Agishtein and A.A.Migdal, 
%Nucl.Phys.B350 (1991) 690; 
%%\bibitem{KKMWIK} 
%H.Kawai, N.Kawamoto, T.Mogami and Y.Watabiki, 
%Phys.Lett.B306 (1993) 19; 
%%
%N.Ishibashi and H.Kawai, 
%Phys.Lett.B314 (1993) 190;
%%
%J.Ambj\o rn, J.Jurkiewicz and Y.Watabiki,
%Nucl.Phys.B454 (1995) 313;
%%
%N.Tsuda and T.Yukawa, 
%Phys.Lett.B305 (1993) 223;
%%
%A.Fujitsu, N.Tsuda and T.Yukawa, hep-lat/9603013, 
%to be published in Int.Jour.Mod.Phys.A. 
%%%%%%%%%%%%%%%%%%%%%%%%%%%%%%%%%%%%%%%%%%%%%%%%%%%%%%%%%%%%%%%%%
\end{thebibliography}
\end{document}